\documentclass[prl, aps, twocolumn, superscriptaddress, 10pt]{revtex4-2}
\usepackage{amssymb}
\usepackage{amsbsy}
\usepackage{amsmath}
\usepackage{amsthm}
\usepackage{graphics}
\usepackage{setspace}
\usepackage{array}
\usepackage{xcolor}
\usepackage{textcomp}
\usepackage{bm}
\usepackage{pifont}
\usepackage[bookmarks=false,linkcolor=blue,urlcolor=blue,colorlinks,citecolor=blue]{hyperref}
\usepackage{soul}
\usepackage[T1]{fontenc}
\usepackage{mathdots}
\usepackage{makecell}
\usepackage{standalone}

\usepackage{dsfont}
\usepackage{bbold}

\DeclareMathOperator{\Tr}{Tr}

\newcommand{\n}{\nonumber}

\newcommand{\blue}[1]{{\color{blue}{#1}}}

\newcommand{\bsub}{\begin{subequations}}
\newcommand{\esub}{\end{subequations}}

\definecolor{darkred}{rgb}{0.8,0,0}
\definecolor{royalblue}{rgb}{0.0, 0.14, 0.4}
\definecolor{magenta}{cmyk}{0,.9,0,0.2}
\definecolor{amethyst}{rgb}{0.6, 0.4, 0.8}
\definecolor{cadmiumgreen}{rgb}{0.0, 0.42, 0.24}
\definecolor{deepcarmine}{rgb}{0.66, 0.13, 0.24}
\definecolor{forestgreen}{rgb}{0.13, 0.55, 0.13}

\newcommand{\beginsupplement}{
        \setcounter{table}{0}
        \renewcommand{\thetable}{S\arabic{table}}
        \setcounter{figure}{0}
        \renewcommand{\thefigure}{S\arabic{figure}}
        \setcounter{equation}{0}
        \renewcommand{\theequation}{S\arabic{equation}}
        \setcounter{section}{0}
        \renewcommand{\thesection}{\Alph{section}}
        \setcounter{subsection}{0}
        \renewcommand{\thesubsection}{\arabic{subsection}}
}

\newcommand{\PHYA}{\affiliation{Department of Physics and Astronomy, Rice University, Houston, Texas 77005, USA}}
\newcommand{\EQMA}{\affiliation{Extreme Quantum Materials Alliance and Smalley-Curl Institute, Rice University, Houston, Texas 77005, USA}}
\newcommand{\RCQM}{\affiliation{Rice Center for Quantum Materials, Rice University, Houston, Texas 77005, USA}}
\newcommand{\RLEMM}{\affiliation{Rice Laboratory for Emergent Magnetic Materials and Smalley-Curl Institute, Rice University, Houston, Texas 77005, USA}}

\begin{document}
\title{Chiral Weyl--Kondo semimetal and circular photogalvanic effect in a prototype Kondo lattice system}

\author{Yuan Fang}
\thanks{These authors contributed equally}
\PHYA 
\EQMA

\author{Kuan-Sen Lin}
\thanks{These authors contributed equally}
\PHYA 
\EQMA

\author{Mounica Mahankali}
\PHYA 
\EQMA

\author{Arushi}
\PHYA 
\RCQM 

\author{Kevin Allen}
\PHYA 
\RCQM 

\author{Sanu Mishra}
\PHYA 
\RCQM 

\author{Emilia Morosan}
\PHYA 
\RCQM
\RLEMM 

\author{Qimiao Si}
\PHYA 
\EQMA
\RCQM
\RLEMM

\begin{abstract}
    Chiral Weyl--Kondo semimetals (cWKSM) provide a setting in which chiral Weyl quasiparticles emerge in the immediate vicinity of the Fermi energy from a Kondo-driven reconstruction of the strongly correlated electronic states in chiral heavy fermion systems (K.-S. Lin et al., arXiv:2602.22185). A defining characteristic of this strongly correlated topological state is the Kramers chiral Weyl fermions in the low-energy quasiparticle states. Recently, experiments in CeGaGe have emerged as a concrete realization of the proposed effect (Arushi et al., preprint). Motivated by these findings, here we go beyond the materials-specific effects by constructing a prototype Kondo lattice model; it incorporates only the essential couplings that respect the associated tetragonal crystalline symmetries. This simplification allows us to robustly demonstrate the symmetry-enforced Kramers Weyl fermions and related topological nodal states in the spectrum of heavy quasiparticles. Furthermore, the simplification provides a tractable setting to determine the salient features in the system's nonlinear optical response, the circular photogalvanic effect, in chiral Weyl--Kondo semimetals. Both analytical and numerical calculations identify sharp peaks in the frequency domain as signatures of the Kondo-driven chiral Weyl nodes; the sharpness of the spectrum reflects the resonant nature of the underlying strongly correlated electronic excitations. Thus, cWKSM provides a unique setting to spectroscopically identify topological fermions that are induced by strong electron correlations. As such, our results are expected to bring about much needed new insights into the understanding of strongly correlated gapless topological matter.
\end{abstract}

\maketitle

\noindent 
\blue{\emph{Introduction}}--- 
Strong electronic correlations can give rise to novel phases of quantum matter~\cite{Keimer2017Physics,Paschen2020Quantum}. Their interplay with topology is particularly intriguing, as it leads to emergent topological states and physical responses that can be drastically different from their noninteracting counterparts.
Heavy fermion systems provide a natural setting for exploring this physics. Through the Kondo effect, local moments acquire electronic quantum numbers, giving rise to strongly renormalized quasiparticles at low energies.
In the presence of nontrivial crystalline symmetry constraints, the emergent heavy-fermion excitations can realize correlated metallic topological phases, the Weyl--Kondo semimetals (WKSMs), in which Weyl nodes emerge in the Kondo-driven low-energy electronic states~\cite{Lai2018Weyl,Grefe2020WeylKondo,Dzsaber2017Kondo,Dzsaber2021Giant,Chen2022Topological,Hu2021Topological,Kirschbaum2026Emergent}. 
Understanding the characteristic responses of these strongly correlated Weyl quasiparticles is essential for identifying and probing Kondo-driven topological systems and thereby for advancing the general understanding of strongly correlated topological matter. 

\begin{figure}[t]
    \centering
    \includegraphics[width=\linewidth]{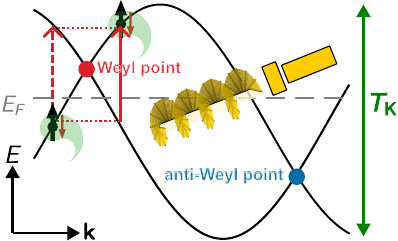}
    \caption{
        Schematic of Weyl and anti-Weyl points formed by heavy fermion quasiparticles sit at different energies~\cite{Lin2026Chiral}. The energy scale of cWKSM is the Kondo energy $k_{\mathrm{B}}T_{\mathrm{K}}$. 
        Shining light excites the quasiparticles out of thermal equilibrium, leading to a net photocurrent.  
        The red dotted line indicates resonant momenta. Solid and dashed red arrow indicate excitations with large and small transitions rates, respectively. 
    }
    \label{fig:scheme}
\end{figure}

This framework has recently been brought to the realm of chiral crystal structures. What results is the chiral Weyl--Kondo semimetal (cWKSM)~\cite{Lin2026Chiral}, which features symmetry-enforced Kramers--Weyl fermions in the spectrum of Kondo-driven quasiparticles pinned to the immediate vicinity of the Fermi energy (see Fig.~\ref{fig:scheme}). Heavy fermions with chiral crystal structures~\cite{Flack2003Chiral,Chang2018Topological,Fecher2022Chirality,Wang2024Chiral,Zyuzin2012Topological,Huang2017Topological} have rarely been studied~\cite{Iwasa2023Weyl}. The materials search strategy outlined in Ref.~\cite{Lin2026Chiral} has identified candidate Weyl--Kondo semimetals, including CePt$_2$B, which are yet to be systematically studied experimentally.

Very recently, CeGaGe has emerged as a candidate material to realize cWKSM~\cite{Arushi2026}. The coexistence of strong correlations, crystal chirality, and Weyl topology raises the question of how the Kondo origin of the low-energy quasiparticles can be revealed in the electronic responses. 
The prospect for measuring these electronic responses in the candidate cWKSM materials adds to the motivation of such exploration.

The chirality of the Weyl nodes affords a topological probe in the form of the circular photogalvanic effect (CPGE), in which circularly polarized light generates a direct photocurrent~\cite{Belinicher1980Photogalvanic,Ivchenko2012Superlattices,Sturman2021Photovoltaic,Ma2021Topology,DeJuan2017Quantized,DeJuan2020Difference,Sipe2000Secondordera,Golub2020Semiclassical,Parker2019Diagrammatic,Morimoto2016Topological,Orenstein2021Topology,Ma2017Direct,Rees2020Helicitydependent,Ni2021Giant}. 
For an ideal noninteracting two-band Weyl node, the trace of the CPGE tensor is quantized to the topological charge of the Weyl node. 
The effect of (weak) electronic interactions, among other factors, can also be systematically treated~\cite{Avdoshkin2020Interactions,Raj2024Photogalvanic,Wu2024Absencea}. 
In a Weyl--Kondo semimetal, the low-energy Weyl states themselves emerge from strong electron correlations, making it particularly intriguing to address how the Kondo effect manifests itself in the CPGE response of a cWKSM. 

In this work, we focus on the strongly correlated topological electronic states of cWKSMs and their signatures in the CPGE. While theoretical studies on the cWKSMs have so far been material specific~\cite{Lin2026Chiral}, here, triggered by the recent experiments on CeGaGe whose low-temperature structure corresponds to space group no.~78~\cite{Arushi2026}, we advance a prototype and, in many ways, the simplest Kondo lattice model for the cWKSM state. 
The model is constructed to respect the chiral crystalline symmetry of space group no.~78 while featuring a very simple electronic dispersion.
This simplified Kondo lattice model on a tetragonal lattice allows us to demonstrate the symmetry-enforced Kramers topological nodes and related Weyl fermions in the spectrum of heavy quasiparticles in a particularly transparent way and readily calculate their chiral charges. In turn, the simplified model affords a setting to uncover the salient features of CPGE associated with the strongly correlated nature of the topological quasiparticles.
We show that the Kondo-driven chiral Weyl nodes manifest through sharp peaks in the frequency domain, which reflect the resonant nature of the Kondo-driven electronic excitations. 
In other words, our work highlights how the cWKSM phase offers a unique setting to identify tell-tale spectroscopic signatures of the strongly correlated gapless topological fermions. 
Because clear-cut signatures of strong correlation-driven gapless topological states are rare, our finding---enabled by the construction of simplified chiral Kondo lattice models---is highly promising for advancing the broad-based field of topology interplaying with strong correlations. 

\begin{figure*}[t]
    \centering
    \includegraphics[width=\linewidth]{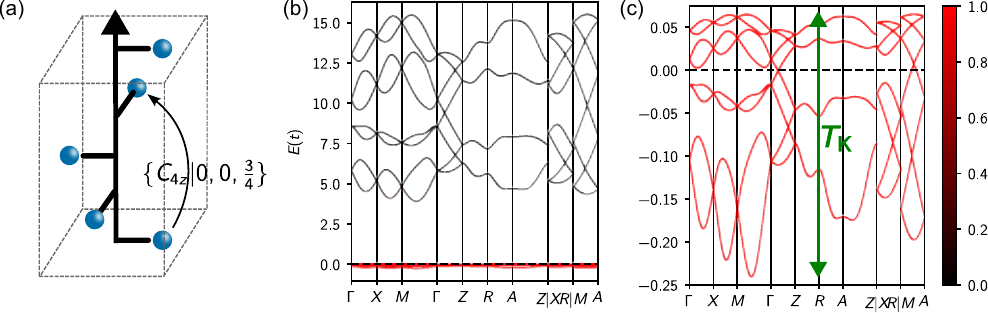}
    \caption{
        (a) Schematic of crystal structure of the prototype model for chiral Weyl--Kondo semimetal in space group no.~78 ($P 4_3$). The dashed gray box indicates the unit cell.  
        (b) Saddle point solution of the Kondo lattice model in the parton representation. Red color indicates $f$ orbital weight in the states. 
        (c) Zoom-in of the $f$-electron dominated narrow bands near the Fermi energy. The bandwidth gives the order of Kondo temperature $T_{\mathrm{K}}$. 
    }
    \label{fig:bs}
\end{figure*}

\noindent 
\blue{\emph{Prototype model}}--- 
We start by constructing a prototype model for cWKSM in space group no.~78 ($P4_3$), which is chiral, noncentrosymmetric, and nonsymmorphic. The space group describes the crystalline symmetries of CeGaGe in the low temperature structure~\cite{Arushi2026}. A schematic of the crystal structure for the prototype model is shown in Fig.~\ref{fig:bs}(a), where the four symmetry-related atom sites are $(x,y,z),(-y,x,z+\frac34),(-x,-y,z+\frac12),(y,-x,z+\frac14)$. 
We place one itinerant conduction electron, labeled as a $c$ electron, and one localized $f$ electron on each site. Each orbital has two spin degrees of freedom. The model contains one type of elementary band representation (EBR)~\cite{cano2021band}. The details of this model can be found in the Supplemental Material (SM)~\cite{sm}. 

To begin with, we construct a simplified noninteracting conduction-electron Hamiltonian $H_0$ that preserves both the crystalline symmetries of space group no.~78 and time-reversal symmetry. 
The model includes the nearest and next-nearest neighbor hoppings and spin-orbit coupling (SOC). 
We then couple the conduction electrons to local, strongly interacting $f$-electron degrees of freedom and consider the resulting periodic Anderson model. The Hamiltonian is
\begin{equation} \label{eqn:H}
\begin{split}
    H = H_0 + \sum_{i,\sigma} \left[ {\epsilon}_f n_{f,i\sigma} + {V} \left( f_{i\sigma}^\dagger c_{i\sigma} + h.c. \right) \right]  \\ 
    -\mu \sum_{i,\sigma} \left( n_{c,i\sigma}+n_{f,i\sigma} \right) + \sum_i U n_{f,i\uparrow} n_{f,i\downarrow}  \,. 
\end{split}
\end{equation}
Here $i$ labels both the unit cell and the sublattice, and $f_{i\sigma}^\dagger$ [$c_{i\sigma}^\dagger$] creates an $f$ [$c$] electron with spin $\sigma$ at site $i$. The corresponding number operators are $n_{f,i\sigma}=f_{i\sigma}^\dagger f_{i\sigma}$ and $n_{c,i\sigma}=c_{i\sigma}^\dagger c_{i\sigma}$. 
We use symbols ${\epsilon}_f$ to denote the bare $f$-electron level, and ${V}$ to denote the bare $c$-$f$ hybridization amplitude. The parameter $\mu$ is the chemical potential, and $U$ is the on-site Coulomb repulsion between the $f$ electrons.

We focus on the strongly correlated regime in which $U$ is much larger than the conduction-electron bandwidth. In the infinite-$U$ limit, double occupation of an $f$ orbital is prohibited, leading to the local constraint $\sum_\sigma \langle n_{f,i\sigma}\rangle \leq 1$.
To implement this constraint, we apply the parton method where an auxiliary boson field $b_i$ and a Lagrange multiplier $\lambda_i$ are introduced~\cite{Hewson1993Kondo}.
At the saddle point level, these fields are replaced by their expectation values, leading to a description in terms of the renormalized $c$-$f$ hybridization and $f$-electron level. The saddle point self-consistent equations and further details of the formulation are given in the SM~\cite{sm}.

The saddle point solution of the prototype model is shown in Fig.~\ref{fig:bs}.
We now describe the symmetry constraints that underlie the rich topological nodes in our prototype model.

\begin{figure}[t]
    \centering
    \includegraphics[width=\linewidth]{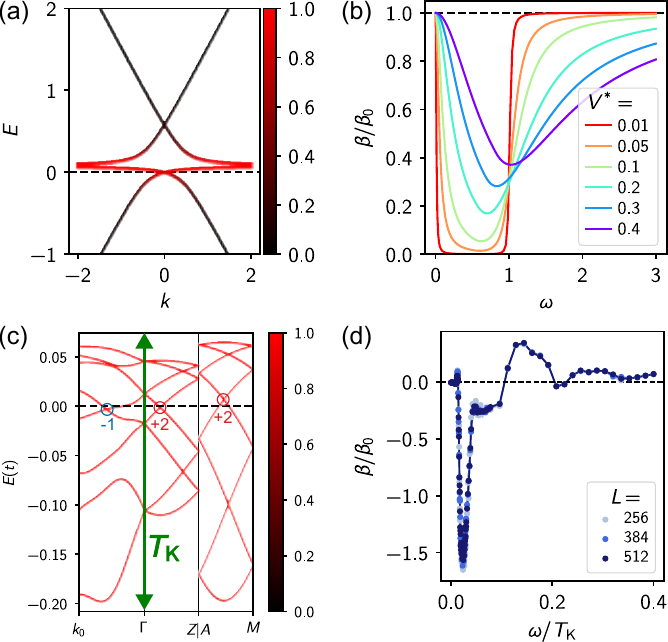}
    \caption{
        (a) Dispersion of a single $f$-electron dominated Weyl node and the $c$-electron dominated Weyl node. Both nodes have the same chiral charge $C=1$. Here we take the renormalized hybridization $V^*=0.2$. 
        (b) The CPGE response $\beta$ as a function of frequency $\omega$ for several hybridization strengths $V^*$.
        (c) The dispersion of the renormalized $f$-electron-dominated Weyl nodes and anti-Weyl nodes near the Fermi energy in our prototype model of space group no.~78 ($P4_3$). The two accordian nodes along $\Gamma-Z$ and $M-A$ have chiral charge $C=2$~\cite{Tsirkin2017Composite,hirschmann2021symmetryenforced} and the accidental Weyl point at general momentum has chiral charge $C=-1$. Here $k_0=(0.372\pi, \pi, 0.572\pi)$. By considering the symmetry-related Weyl points from the marked Weyl points in (c), the summation of chiral charges is zero, satisfying the Nielsen--Ninomiya theorem~\cite{Nielsen1981Absence,Nielsen1981Absencea}.
        (d) The CPGE response $\beta$ as a function of frequency $\omega/T_{\mathrm{K}}$, calculated on the lattice of size $L\times L\times L$. 
    }
    \label{fig:CPGE}
\end{figure}

\noindent 
\blue{\emph{Symmetry enforced Weyl nodes and their chiral charges}}--- 
In the presence of SOC and time-reversal ($\mathcal{T}$) symmetry, the quasiparticle excitation spectrum of our prototype model with space group no.~78 ($P4_3$) exhibits rich topological nodes enforced by crystalline symmetries~\cite{hirschmann2021symmetryenforced}, as demonstrated in Fig.~\ref{fig:bs}.
First, along the high-symmetry lines $\Gamma-Z$ and $M-A$, the screw symmetry $\{ C_{4z} | 0,0,\frac{3}{4}\}$ [Fig.~\ref{fig:bs}(a)] enforces the accordion-type band connectivity.
Second, along the high-symmetry line $X-R$, the screw symmetry $\{ C_{2z} | 0,0,\frac{1}{2}\}$ enforces the hourglass-type band connectivity.
The band crossings occur along $\Gamma-Z$, $M-A$, $X-R$ are symmetry-enforced Weyl points.
Third, due to the composite symmetry $\{ C_{2z} \mathcal{T} | 0,0,\frac{1}{2} \}$, the quasiparticle excitation spectrum has an enforced twofold degeneracy on the $k_z = \pi$ plane, which corresponds to a Weyl nodal plane.
Finally, the structural chirality features Weyl points at the time-reversal-invariant momenta (TRIMs), which are the so-called Kramers--Weyl points~\cite{Chang2018Topological}.
In our prototype model, the Kramers--Weyl points occur at $\Gamma$, $X$, and $M$.
We numerically evaluate the chiral charges~\cite{Armitage2018Weyl} for some of the aforementioned Weyl points in our prototype model and label them in Fig.~\ref{fig:CPGE}(c).
We note that, in addition to the symmetry-enforced Weyl nodes, there are also Weyl nodes at general momenta, as indicated in Fig.~\ref{fig:CPGE}(c).
We also numerically evaluate the chiral charges of Kramers--Weyl points in the $f$-electron-dominated bands, which take values of $+1$ and $-1$.
Importantly, due to the structural chirality, the Weyl points that carry opposite signs of chiral charges are not located at the same energy, where Fig.~\ref{fig:CPGE}(c) is a demonstration.
Hence, our prototype model provides an ideal platform to investigate the influence of these heavy-fermion topological nodes on the CPGE responses.
Further details of the symmetry-enforced band crossings and chiral charges are 
provided in the SM~\cite{sm}.

\noindent 
\blue{\emph{CPGE in cWKSM}}--- 
CPGE is the part of photocurrent that is anti-symmetric in circular polarization of the incident light, and is an established optical spectroscopy to probe the electronic excitations in condensed matter~\cite{Belinicher1980Photogalvanic,Ivchenko2012Superlattices,Sturman2021Photovoltaic}. 
In chiral materials, the CPGE tensor is defined as the second order response $\frac{dj^a}{dt} = \beta^{aa'} \left( \mathbf E(\omega) \times  \mathbf E^* (-\omega)\right)_{a'}$~\cite{DeJuan2017Quantized}, where $a,a'$ are spatial indices.
For an ideal noninteracting Weyl semimetal, the CPGE exhibits a quantized response over a finite frequency window in which optical transitions involve only a single Weyl node. In this regime, the trace of the CPGE tensor is determined solely by the topological charge $C$ of the Weyl node: $\beta \equiv -i\mathrm{Tr}\,\beta_{ab} = C\beta_0$ where $\beta_0 = \frac{\pi e^3}{h^2}$. 
This quantization, however, relies on the ideal two-band description and, in realistic systems, will be affected by the multiplicity of the bands and electron interactions~\cite{Avdoshkin2020Interactions,Raj2024Photogalvanic,Wu2024Absencea}.
Here we study what happens in the Kondo systems.
As we will show, our key finding is that the Kondo-driven chiral Weyl nodes lead to characteristic resonance-like sharp features of the CPGE spectrum in the frequency domain.

To understand the CPGE response in the presence of the Kondo effect, we first consider a single $f$-electron-dominated Weyl node generated by the hybridization between $f$ and $c$ electrons. The noninteracting conduction-electron Hamiltonian near the single Weyl node is given by
\begin{equation}\label{eqn:Weyl_Hamiltonian}
    H_0=\sum_{\mathbf k,\alpha,\alpha'}c^\dagger_{\mathbf k,\alpha}\left[v_{\mathrm{F}}(\mathbf k+\mathbf A)\cdot\vec{\sigma}_{\alpha\alpha'}-\mu\delta_{\alpha\alpha'}\right]c_{\mathbf k,\alpha'},
\end{equation}
where $\alpha, \alpha'$ label spin and the conduction electrons couple to electromagnetic field through the vector potential $\mathbf A$. At the saddle-point level, the hybridization with the $f$ electrons introduces a frequency-dependent self-energy to the conduction-electron Green's function, $G_c^{-1}(\mathbf k,\omega) = \omega - \xi v_{\mathrm{F}}\mathbf k\cdot\vec{\sigma}+\mu-\Sigma(\omega)$ where $\Sigma(\omega)= \frac{V^{*2}}{\omega-\epsilon_f^*}$ and $\xi=\pm$ labels the left/right-propagating branches. 
Here $V^*$ is the renormalized hybridization and ${\epsilon}_f^*$ is the renormalized $f$-electron level. Therefore, although the electromagnetic field couples directly to the conduction electrons, the optical response is determined by the fully dressed $c$-electron propagators containing the Kondo hybridization.

The CPGE is a second-order optical response and, within our
saddle-point approach, is given diagrammatically by triangle diagrams constructed from the full $c$-electron propagators and the electromagnetic vertices. Equivalently, since the saddle-point problem is quadratic, the same result can be expressed in terms of the renormalized quasiparticle bands obtained from the poles of the full Green's function. The resulting CPGE coefficient $\beta^{aa'}$ is compactly written in the Sipe--Shkrebtii form~\cite{Sipe2000Secondordera}
\begin{equation}
    \beta^{aa'}=\pi\epsilon^{a'bc}\int_{\mathbf k}\sum_{nm}\frac{\partial\epsilon_{nm}}{\partial k_a}r^b_{nm}r^c_{mn}f_{nm}\delta(\omega-\epsilon_{nm}),
\end{equation}
where $\int_{\mathbf{k}}\equiv \int \frac{d^3 k}{(2\pi)^3}$, $\epsilon_{nm}=\epsilon_n(\mathbf{k})-\epsilon_m(\mathbf{k})$, $f_{nm}=f_n-f_m$, and $r^b_{nm}=i\langle u_n(\mathbf{k})|\partial_{k_b}|u_m(\mathbf{k})\rangle$. Here $\epsilon_n(\mathbf{k})$ and $|u_n(\mathbf{k})\rangle$ are the energies and the periodic part of Bloch states, $f_n$ is the Fermi--Dirac distribution function for the $n$-th band. 
The delta function restricts the response to interband transitions that satisfy the resonance condition $\omega=\epsilon_{nm}$. Thus, a given photon frequency $\omega$ selects a resonant momentum $\mathbf{k}^*$, and the CPGE response is determined by the optical matrix elements of the transition at that momentum as illustrated in Fig.~\ref{fig:scheme}. 

For the single-Weyl-node model, the quasiparticle energies and wave functions can both be obtained analytically. The four hybridized bands are
$\epsilon_{\xi\eta}(k)=\frac{\xi v_{\mathrm{F}}k+{\epsilon}_f^*}{2}+\eta\sqrt{\left(\frac{\xi v_{\mathrm{F}}k-{\epsilon}_f^*}{2}\right)^2+V^{*2}}$
where $\xi=\pm$ labels the left- and right- propagating branches and $\eta=\pm$ labels the upper and lower bands. The saddle-point quasiparticle weight is characterized by $Z_{\xi\eta}^{-1} = 1-\partial_z\Sigma(z) \big|_{z=\epsilon_{\xi\eta}}$ which gives rise to:
\begin{equation}
    Z_{\xi\eta}^{-1}=1+\frac{V^{*2}}{\left(\epsilon_{\xi\eta}-{\epsilon}_f^*\right)^2}.
\end{equation}
This is equal to the conduction-electron component of each quasiparticle wave function. 
Since the electromagnetic field couples to the $c$-electron sector, these conduction-electron weights directly enter the optical matrix elements. 
The conventional definition of the parton quasiparticle weight is related to $Z_{\xi\eta}$ (see SM~\cite{sm}).

In this approach, the CPGE response of the single Weyl node is a direct consequence of chiral anomaly~\cite{Adler1969AxialVector,Bell1969PCAC,Adler1969Absence}, and can be evaluated explicitly as~\cite{sm}
\begin{equation}\label{eqn:beta}
    \beta(\omega)=\beta_0\sum_{\eta=\pm}Z_{--}Z_{+\eta}\left(\frac{2v_{\mathrm{F}}k^*_\eta}{\omega}\right)^2\Theta(k^*_\eta-k_{\mathrm{F}}),
\end{equation}
where $k_{\mathrm{F}}$ is the Fermi momentum and $k^*_\eta$ is the renormalized resonant momentum determined by the resonance condition
$\omega=\epsilon_{+\eta}(k^*_\eta)-\epsilon_{--}(k^*_\eta)$.
The final expression contains three corrections arising from the Kondo hybridization. 
The first two factors, $Z_{--}$ and $Z_{+\eta}$, are the conduction-electron weights of the two bands participating in the resonant optical transition. The third factor, $(2k^*_\eta/\omega)^2$, originates from the renormalization of the quasiparticle dispersion, which modifies the relation between the resonant momentum $k^*_\eta$ and the photon frequency $\omega$. 
These three factors together determine how the Kondo effect modifies the CPGE response. 
A model with opposite chirality with $C=-1$ will reverse the sign of the overall contribution to Eq.~\eqref{eqn:beta}. 

Figure~\ref{fig:CPGE}(a) shows the saddle-point solution of the single-Weyl-node model. The corresponding CPGE response $\beta$ as a function of frequency $\omega$ is shown in Fig.~\ref{fig:CPGE}(b) for several renormalized hybridization strengths $V^*$, while the renormalized $f$-electron level ${\epsilon}_f^*$ and the chemical potential $\mu$ are kept fixed. 
In this model, the CPGE coefficient deviates from the noninteracting plateau $\beta/\beta_0=1$ at frequencies smaller than twice the $c$-Weyl energy level. 
In this Kondo regime, a peak shows near $\omega=0$ and then it quickly decays due to the strong renormalization of heavy fermion quasiparticles as shown in Fig.~\ref{fig:CPGE}(b), thereby capturing the resonance nature of the Kondo-driven electronic excitations.
When $\omega \gg T_{\mathrm{K}}$, it recovers the noninteracting plateau where the Kondo hybridization effect is negligible.

We then numerically compute the CPGE response for the prototype model in Eq.~\eqref{eqn:H}, which contains multiple Weyl nodes. The dispersion of the $f$-electron-dominated Weyl nodes near the Fermi energy is shown in Fig.~\ref{fig:CPGE}(c). The calculated CPGE response $\beta$ as a function of the scaled frequency $\omega/T_{\mathrm K}$ is presented in Fig.~\ref{fig:CPGE}(d).
Due to the Kondo hybridization, the CPGE response is peaked at the frequency corresponding to the energy levels of the heavy fermion Weyl nodes, and is reduced at other frequencies.

\noindent 
\blue{\emph{Conclusion and discussion}}--- 
In this work, we studied cWKSM by constructing a prototype Kondo lattice model that captures the space group symmetry of the recently discovered material candidate CeGaGe~\cite{Arushi2026} but involves the simplest couplings.
We obtained $f$-electron-dominated Weyl quasiparticles near the Fermi energy and determined their topological charges. 

Within this prototype model, we are able to calculate the CPGE response.
We showed that the Kondo-driven Weyl nodes enter the CPGE through the quasiparticle renormalization; more specifically, 
the Kondo effect appears through 
the conduction-electron weights of the electronic states involved in the resonant transition, as well as by the change of the resonant momentum due to the strong renormalization effect. What emerges is a characteristic sharp feature in the CPGE spectrum: Because of the chirality enabled separation in energy between the Kondo-driven Weyl and anti-Weyl nodes, the CPGE spectrum displays Kondo-like resonance peaks in the frequency domain.

Accordingly, based on cWKSM, our results uncover a tell-tale sign of Weyl nodes that inherently features strong correlations.
The results provide a rare spectroscopic means of directly implicating the strongly correlated nature of gapless topological fermions. As such, we can expect our finding to shed much new light into the understanding of strongly correlated topological metals.

\noindent 
\blue{\emph{Acknowledgment}}--- 
We would like to thank Fang Xie, Chenyuan Li, Joel Moore and Silke Paschen for useful discussions. 
The work has primarily been supported by the NSF Grant No.\ DMR-2220603 (YF, KSL),
the Robert A. Welch Foundation Grant No.\ C-1411 (MM) and  the Vannevar Bush Faculty Fellowship ONR-VB N00014-23-1-2870 (MM, QS). 
K.-S.L. acknowledges the Carl and Lillian Illig Postdoctoral Fellowship from the Smalley-Curl Institute at Rice University.
The work has in part been supported by
the Vannevar Bush Faculty Fellowship ONR-VB N00014-24-1-2048 (A, HB, EM) and the Robert A. Welch Foundation Grant No. C-2114 (SM, EM).
QS and EM acknowledge support from the DOE, BES Grant No.\ DE-SC0026179.
The computational calculations have in part been performed on the Shared University Grid at Rice funded by NSF under Grant EIA-0216467, 
a partnership between Rice University, Sun Microsystems, and Sigma Solutions, Inc., the Big-Data Private-Cloud Research Cyberinfrastructure
MRI-award funded by NSF under Grant No. CNS-1338099, and the Extreme Science and Engineering Discovery Environment (XSEDE) by NSF under Grant No. DMR170109. 

\bibliography{reference.bib}

@misc{sm,
  note = {See Supplemental Material.}
}

@article{Arushi2026,
  journal = {to appear},
  author = {Arushi et al., A },
  year = 2026,
}

@article{cano2021band,
author = {Cano, Jennifer and Bradlyn, Barry},
title = {Band Representations and Topological Quantum Chemistry},
journal = {Annual Review of Condensed Matter Physics},
volume = {12},
number = {1},
pages = {225-246},
year = {2021},
doi = {10.1146/annurev-conmatphys-041720-124134},
URL = { https://doi.org/10.1146/annurev-conmatphys-041720-124134
},
eprint = { https://doi.org/10.1146/annurev-conmatphys-041720-124134}
}

@article{Armitage2018Weyl,
  title = {Weyl and {{Dirac}} Semimetals in Three-Dimensional Solids},
  author = {Armitage, N. P. and Mele, E. J. and Vishwanath, Ashvin},
  year = 2018,
  month = jan,
  journal = {Reviews of Modern Physics},
  volume = {90},
  number = {1},
  pages = {015001},
  issn = {0034-6861, 1539-0756},
  doi = {10.1103/RevModPhys.90.015001},
  url = {https://link.aps.org/doi/10.1103/RevModPhys.90.015001},
  langid = {english}
}

@article{Avdoshkin2020Interactions,
  title = {Interactions {{Remove}} the {{Quantization}} of the {{Chiral Photocurrent}} at {{Weyl Points}}},
  author = {Avdoshkin, Alexander and Kozii, Vladyslav and Moore, Joel E.},
  year = 2020,
  month = may,
  journal = {Physical Review Letters},
  volume = {124},
  number = {19},
  pages = {196603},
  issn = {0031-9007, 1079-7114},
  doi = {10.1103/PhysRevLett.124.196603},
  url = {https://link.aps.org/doi/10.1103/PhysRevLett.124.196603},
  langid = {english}
}

@article{Chang2018Topological,
  title = {Topological Quantum Properties of Chiral Crystals},
  author = {Chang, Guoqing and Wieder, Benjamin J. and Schindler, Frank and Sanchez, Daniel S. and Belopolski, Ilya and Huang, Shin-Ming and Singh, Bahadur and Wu, Di and Chang, Tay-Rong and Neupert, Titus and Xu, Su-Yang and Lin, Hsin and Hasan, M. Zahid},
  year = 2018,
  month = nov,
  journal = {Nature Materials},
  volume = {17},
  number = {11},
  pages = {978--985},
  issn = {1476-1122, 1476-4660},
  doi = {10.1038/s41563-018-0169-3},
  url = {https://www.nature.com/articles/s41563-018-0169-3},
  langid = {english}
}

@article{Chen2022Topological,
  title = {Topological Semimetal Driven by Strong Correlations and Crystalline Symmetry},
  author = {Chen, Lei and Setty, Chandan and Hu, Haoyu and Vergniory, Maia G. and Grefe, Sarah E. and Fischer, Lukas and Yan, Xinlin and Eguchi, Gaku and Prokofiev, Andrey and Paschen, Silke and Cano, Jennifer and Si, Qimiao},
  year = 2022,
  month = nov,
  journal = {Nature Physics},
  volume = {18},
  number = {11},
  pages = {1341--1346},
  issn = {1745-2473, 1745-2481},
  doi = {10.1038/s41567-022-01743-4},
  url = {https://www.nature.com/articles/s41567-022-01743-4},
  langid = {english}
}

@article{DeJuan2017Quantized,
  title = {Quantized Circular Photogalvanic Effect in {{Weyl}} Semimetals},
  author = {De Juan, Fernando and Grushin, Adolfo G. and Morimoto, Takahiro and Moore, Joel E},
  year = 2017,
  month = jul,
  journal = {Nature Communications},
  volume = {8},
  number = {1},
  pages = {15995},
  issn = {2041-1723},
  doi = {10.1038/ncomms15995},
  url = {https://www.nature.com/articles/ncomms15995},
  langid = {english}
}

@article{DeJuan2020Difference,
  title = {Difference Frequency Generation in Topological Semimetals},
  author = {De Juan, F. and Zhang, Y. and Morimoto, T. and Sun, Y. and Moore, J. E. and Grushin, A. G.},
  year = 2020,
  month = jan,
  journal = {Physical Review Research},
  volume = {2},
  number = {1},
  pages = {012017},
  issn = {2643-1564},
  doi = {10.1103/PhysRevResearch.2.012017},
  url = {https://link.aps.org/doi/10.1103/PhysRevResearch.2.012017},
  langid = {english}
}

@article{Dzsaber2017Kondo,
  title = {Kondo {{Insulator}} to {{Semimetal Transformation Tuned}} by {{Spin-Orbit Coupling}}},
  author = {Dzsaber, S. and Prochaska, L. and Sidorenko, A. and Eguchi, G. and Svagera, R. and Waas, M. and Prokofiev, A. and Si, Q. and Paschen, S.},
  year = 2017,
  month = jun,
  journal = {Physical Review Letters},
  volume = {118},
  number = {24},
  pages = {246601},
  issn = {0031-9007, 1079-7114},
  doi = {10.1103/PhysRevLett.118.246601},
  url = {http://link.aps.org/doi/10.1103/PhysRevLett.118.246601},
  copyright = {http://link.aps.org/licenses/aps-default-license},
  langid = {english}
}

@article{Dzsaber2021Giant,
  title = {Giant Spontaneous {{Hall}} Effect in a Nonmagnetic {{Weyl}}--{{Kondo}} Semimetal},
  author = {Dzsaber, Sami and Yan, Xinlin and Taupin, Mathieu and Eguchi, Gaku and Prokofiev, Andrey and Shiroka, Toni and Blaha, Peter and Rubel, Oleg and Grefe, Sarah E. and Lai, Hsin-Hua and Si, Qimiao and Paschen, Silke},
  year = 2021,
  month = feb,
  journal = {Proceedings of the National Academy of Sciences},
  volume = {118},
  number = {8},
  pages = {e2013386118},
  issn = {0027-8424, 1091-6490},
  doi = {10.1073/pnas.2013386118},
  url = {https://pnas.org/doi/full/10.1073/pnas.2013386118},
  langid = {english}
}

@article{Fecher2022Chirality,
  title = {Chirality in the {{Solid State}}: {{Chiral Crystal Structures}} in {{Chiral}} and {{Achiral Space Groups}}},
  shorttitle = {Chirality in the {{Solid State}}},
  author = {Fecher, Gerhard H. and K{\"u}bler, J{\"u}rgen and Felser, Claudia},
  year = 2022,
  month = aug,
  journal = {Materials},
  volume = {15},
  number = {17},
  pages = {5812},
  issn = {1996-1944},
  doi = {10.3390/ma15175812},
  url = {https://www.mdpi.com/1996-1944/15/17/5812},
  langid = {english}
}

@article{Flack2003Chiral,
  title = {Chiral and {{Achiral Crystal Structures}}},
  author = {Flack, Howard D.},
  year = 2003,
  month = apr,
  journal = {Helvetica Chimica Acta},
  volume = {86},
  number = {4},
  pages = {905--921},
  issn = {0018-019X, 1522-2675},
  doi = {10.1002/hlca.200390109},
  url = {https://onlinelibrary.wiley.com/doi/10.1002/hlca.200390109},
  copyright = {http://onlinelibrary.wiley.com/termsAndConditions\#vor},
  langid = {english}
}

@article{Golub2020Semiclassical,
  title = {Semiclassical Theory of the Circular Photogalvanic Effect in Gyrotropic Systems},
  author = {Golub, L. E. and Ivchenko, E.~L. and Spivak, B.},
  year = 2020,
  month = aug,
  journal = {Physical Review B},
  volume = {102},
  number = {8},
  pages = {085202},
  issn = {2469-9950, 2469-9969},
  doi = {10.1103/PhysRevB.102.085202},
  url = {https://link.aps.org/doi/10.1103/PhysRevB.102.085202},
  langid = {english}
}

@article{Grefe2020WeylKondo,
  title = {Weyl-{{Kondo}} Semimetals in Nonsymmorphic Systems},
  author = {Grefe, Sarah E. and Lai, Hsin-Hua and Paschen, Silke and Si, Qimiao},
  year = 2020,
  month = feb,
  journal = {Physical Review B},
  volume = {101},
  number = {7},
  pages = {075138},
  issn = {2469-9950, 2469-9969},
  doi = {10.1103/PhysRevB.101.075138},
  url = {https://link.aps.org/doi/10.1103/PhysRevB.101.075138},
  langid = {english}
}

@book{Hewson1993Kondo,
  title = {The {{Kondo Problem}} to {{Heavy Fermions}}},
  author = {Hewson, Alexander Cyril},
  year = 1993,
  month = jan,
  edition = {1},
  publisher = {Cambridge University Press},
  doi = {10.1017/CBO9780511470752},
  url = {https://www.cambridge.org/core/product/identifier/9780511470752/type/book},
  copyright = {https://www.cambridge.org/core/terms},
  isbn = {978-0-521-36382-2 978-0-521-59947-4 978-0-511-47075-2}
}

@article{hirschmann2021symmetryenforced,
  title = {Symmetry-enforced band crossings in tetragonal materials: Dirac and Weyl degeneracies on points, lines, and planes},
  author = {Hirschmann, Moritz M. and Leonhardt, Andreas and Kilic, Berkay and Fabini, Douglas H. and Schnyder, Andreas P.},
  journal = {Phys. Rev. Mater.},
  volume = {5},
  issue = {5},
  pages = {054202},
  numpages = {28},
  year = {2021},
  month = {May},
  publisher = {American Physical Society},
  doi = {10.1103/PhysRevMaterials.5.054202},
  url = {https://link.aps.org/doi/10.1103/PhysRevMaterials.5.054202}
}

@article{Huang2017Topological,
  title = {Topological Responses from Chiral Anomaly in Multi-{{Weyl}} Semimetals},
  author = {Huang, Ze-Min and Zhou, Jianhui and Shen, Shun-Qing},
  year = 2017,
  month = aug,
  journal = {Physical Review B},
  volume = {96},
  number = {8},
  eprint = {1705.04576},
  primaryclass = {cond-mat.mes-hall},
  pages = {085201},
  issn = {2469-9950, 2469-9969},
  doi = {10.1103/PhysRevB.96.085201},
  url = {http://arxiv.org/abs/1705.04576},
  archiveprefix = {arXiv}
}

@article{Iwasa2023Weyl,
  title = {Weyl--{{Kondo}} Semimetal Behavior in the Chiral Structure Phase of {{Ce}} 3 {{Rh}} 4 {{Sn}} 13},
  author = {Iwasa, Kazuaki and Suyama, Kazuya and {Ohira-Kawamura}, Seiko and Nakajima, Kenji and Raymond, St{\'e}phane and Steffens, Paul and Yamada, Akira and Matsuda, Tatsuma D. and Aoki, Yuji and Kawasaki, Ikuto and Fujimori, Shin-ichi and Yamagami, Hiroshi and Yokoyama, Makoto},
  year = 2023,
  month = jan,
  journal = {Physical Review Materials},
  volume = {7},
  number = {1},
  pages = {014201},
  issn = {2475-9953},
  doi = {10.1103/PhysRevMaterials.7.014201},
  url = {https://link.aps.org/doi/10.1103/PhysRevMaterials.7.014201},
  langid = {english}
}

@article{Keimer2017Physics,
  title = {The Physics of Quantum Materials},
  author = {Keimer, B. and Moore, J. E.},
  year = 2017,
  month = nov,
  journal = {Nature Physics},
  volume = {13},
  number = {11},
  pages = {1045--1055},
  issn = {1745-2473, 1745-2481},
  doi = {10.1038/nphys4302},
  url = {https://www.nature.com/articles/nphys4302},
  langid = {english}
}

@article{Lai2018Weyl,
  title = {Weyl--{{Kondo}} Semimetal in Heavy-Fermion Systems},
  author = {Lai, Hsin-Hua and Grefe, Sarah E. and Paschen, Silke and Si, Qimiao},
  year = 2018,
  month = jan,
  journal = {Proceedings of the National Academy of Sciences},
  volume = {115},
  number = {1},
  pages = {93--97},
  issn = {0027-8424, 1091-6490},
  doi = {10.1073/pnas.1715851115},
  url = {https://pnas.org/doi/full/10.1073/pnas.1715851115},
  langid = {english}
}

@article{Lin2026Chiral,
       author = {{Lin}, Kuan-Sen and {Fang}, Yuan and {Fabrelli}, Henrique and {Li}, Runhan and {Prokofiev}, Andrey and {Xie}, Fang and {Cano}, Jennifer and {Vergniory}, Maia G. and {Paschen}, Silke and {Si}, Qimiao},
        title = "{Chiral Weyl-Kondo semimetals and hexagonal heavy fermion systems}",
      journal = {arXiv e-prints},
         year = 2026,
        month = feb,
          eid = {arXiv:2602.22185},
        pages = {arXiv:2602.22185},
          doi = {10.48550/arXiv.2602.22185},
archivePrefix = {arXiv},
       eprint = {2602.22185},
 primaryClass = {cond-mat.str-el}
}

@article{Morimoto2016Topological,
  title = {Topological Nature of Nonlinear Optical Effects in Solids},
  author = {Morimoto, Takahiro and Nagaosa, Naoto},
  year = 2016,
  month = may,
  journal = {Science Advances},
  volume = {2},
  number = {5},
  pages = {e1501524},
  issn = {2375-2548},
  doi = {10.1126/sciadv.1501524},
  url = {https://www.science.org/doi/10.1126/sciadv.1501524},
  langid = {english}
}

@article{Nielsen1981Absence,
  title = {Absence of Neutrinos on a Lattice},
  author = {Nielsen, H.B. and Ninomiya, M.},
  year = 1981,
  month = jul,
  journal = {Nuclear Physics B},
  volume = {185},
  number = {1},
  pages = {20--40},
  issn = {05503213},
  doi = {10.1016/0550-3213(81)90361-8},
  url = {https://linkinghub.elsevier.com/retrieve/pii/0550321381903618},
  copyright = {https://www.elsevier.com/tdm/userlicense/1.0/},
  langid = {english}
}

@article{Nielsen1981Absencea,
  title = {Absence of Neutrinos on a Lattice},
  author = {Nielsen, H.B. and Ninomiya, M.},
  year = 1981,
  month = dec,
  journal = {Nuclear Physics B},
  volume = {193},
  number = {1},
  pages = {173--194},
  issn = {05503213},
  doi = {10.1016/0550-3213(81)90524-1},
  url = {https://linkinghub.elsevier.com/retrieve/pii/0550321381905241},
  copyright = {https://www.elsevier.com/tdm/userlicense/1.0/},
  langid = {english}
}

@article{Orenstein2021Topology,
  title = {Topology and {{Symmetry}} of {{Quantum Materials}} via {{Nonlinear Optical Responses}}},
  author = {Orenstein, J. and Moore, J.E. and Morimoto, T. and Torchinsky, D.H. and Harter, J.W. and Hsieh, D.},
  year = 2021,
  month = mar,
  journal = {Annual Review of Condensed Matter Physics},
  volume = {12},
  number = {1},
  pages = {247--272},
  issn = {1947-5454, 1947-5462},
  doi = {10.1146/annurev-conmatphys-031218-013712},
  url = {https://www.annualreviews.org/doi/10.1146/annurev-conmatphys-031218-013712},
  langid = {english}
}

@article{Parker2019Diagrammatic,
  title = {Diagrammatic Approach to Nonlinear Optical Response with Application to {{Weyl}} Semimetals},
  author = {Parker, Daniel E. and Morimoto, Takahiro and Orenstein, Joseph and Moore, Joel E.},
  year = 2019,
  month = jan,
  journal = {Physical Review B},
  volume = {99},
  number = {4},
  pages = {045121},
  publisher = {American Physical Society},
  doi = {10.1103/PhysRevB.99.045121},
  url = {https://link.aps.org/doi/10.1103/PhysRevB.99.045121}
}

@article{Paschen2020Quantum,
  title = {Quantum Phases Driven by Strong Correlations},
  author = {Paschen, Silke and Si, Qimiao},
  year = 2020,
  month = dec,
  journal = {Nature Reviews Physics},
  volume = {3},
  number = {1},
  pages = {9--26},
  issn = {2522-5820},
  doi = {10.1038/s42254-020-00262-6},
  url = {https://www.nature.com/articles/s42254-020-00262-6},
  langid = {english}
}

@article{Raj2024Photogalvanic,
  title = {Photogalvanic Response in Multi-{{Weyl}} Semimetals},
  author = {Raj, Arpit and Chaudhary, Swati and Fiete, Gregory A.},
  year = 2024,
  month = jan,
  journal = {Physical Review Research},
  volume = {6},
  number = {1},
  pages = {013048},
  issn = {2643-1564},
  doi = {10.1103/PhysRevResearch.6.013048},
  url = {https://link.aps.org/doi/10.1103/PhysRevResearch.6.013048},
  langid = {english}
}

@article{Sipe2000Secondordera,
  title = {Second-Order Optical Response in Semiconductors},
  author = {Sipe, J. E. and Shkrebtii, A. I.},
  year = 2000,
  month = feb,
  journal = {Physical Review B},
  volume = {61},
  number = {8},
  pages = {5337--5352},
  issn = {0163-1829, 1095-3795},
  doi = {10.1103/PhysRevB.61.5337},
  url = {https://link.aps.org/doi/10.1103/PhysRevB.61.5337},
  copyright = {http://link.aps.org/licenses/aps-default-license},
  langid = {english}
}

@article{Tsirkin2017Composite,
  title = {Composite {{Weyl}} Nodes Stabilized by Screw Symmetry with and without Time-Reversal Invariance},
  author = {Tsirkin, Stepan S. and Souza, Ivo and Vanderbilt, David},
  year = 2017,
  month = jul,
  journal = {Physical Review B},
  volume = {96},
  number = {4},
  pages = {045102},
  issn = {2469-9950, 2469-9969},
  doi = {10.1103/PhysRevB.96.045102},
  url = {http://link.aps.org/doi/10.1103/PhysRevB.96.045102},
  copyright = {http://link.aps.org/licenses/aps-default-license},
  langid = {english}
}

@article{Wang2024Chiral,
  title = {Chiral {{Quantum Materials}}: {{When Chemistry Meets Physics}}},
  shorttitle = {Chiral {{Quantum Materials}}},
  author = {Wang, Xia and Yi, Changjiang and Felser, Claudia},
  year = 2024,
  month = mar,
  journal = {Advanced Materials},
  volume = {36},
  number = {13},
  pages = {2308746},
  issn = {0935-9648, 1521-4095},
  doi = {10.1002/adma.202308746},
  url = {https://advanced.onlinelibrary.wiley.com/doi/10.1002/adma.202308746},
  langid = {english}
}

@article{Wu2024Absencea,
  title = {Absence of Quantization in the Circular Photogalvanic Effect in Disordered Chiral {{Weyl}} Semimetals},
  author = {Wu, Ang-Kun and Guerci, Daniele and Fu, Yixing and Wilson, Justin H. and Pixley, J. H.},
  year = 2024,
  month = jul,
  journal = {Physical Review B},
  volume = {110},
  number = {1},
  pages = {014201},
  issn = {2469-9950, 2469-9969},
  doi = {10.1103/PhysRevB.110.014201},
  url = {https://link.aps.org/doi/10.1103/PhysRevB.110.014201},
  langid = {english}
}

@article{Zyuzin2012Topological,
  title = {Topological Response in {{Weyl}} Semimetals and the Chiral Anomaly},
  author = {Zyuzin, A. A. and Burkov, A. A.},
  year = 2012,
  month = sep,
  journal = {Physical Review B},
  volume = {86},
  number = {11},
  pages = {115133},
  issn = {1098-0121, 1550-235X},
  doi = {10.1103/PhysRevB.86.115133},
  url = {https://link.aps.org/doi/10.1103/PhysRevB.86.115133},
  copyright = {http://link.aps.org/licenses/aps-default-license},
  langid = {english}
}

@article{Adler1969Absence,
  title = {Absence of {{Higher-Order Corrections}} in the {{Anomalous Axial-Vector Divergence Equation}}},
  author = {Adler, Stephen L. and Bardeen, William A.},
  year = 1969,
  month = jun,
  journal = {Physical Review},
  volume = {182},
  number = {5},
  pages = {1517--1536},
  issn = {0031-899X},
  doi = {10.1103/PhysRev.182.1517},
  url = {https://link.aps.org/doi/10.1103/PhysRev.182.1517},
  copyright = {http://link.aps.org/licenses/aps-default-license},
  langid = {english}
}

@article{Adler1969AxialVector,
  title = {Axial-{{Vector Vertex}} in {{Spinor Electrodynamics}}},
  author = {Adler, Stephen L.},
  year = 1969,
  month = jan,
  journal = {Physical Review},
  volume = {177},
  number = {5},
  pages = {2426--2438},
  issn = {0031-899X},
  doi = {10.1103/PhysRev.177.2426},
  url = {https://link.aps.org/doi/10.1103/PhysRev.177.2426},
  copyright = {http://link.aps.org/licenses/aps-default-license},
  langid = {english}
}

@article{Bell1969PCAC,
  title = {A {{PCAC}} Puzzle: {$\pi$}0{$\rightarrow\gamma\gamma$} in the {$\sigma$}-Model},
  shorttitle = {A {{PCAC}} Puzzle},
  author = {Bell, J. S. and Jackiw, R.},
  year = 1969,
  month = mar,
  journal = {Il Nuovo Cimento A},
  volume = {60},
  number = {1},
  pages = {47--61},
  issn = {0369-3546, 1826-9869},
  doi = {10.1007/BF02823296},
  url = {http://link.springer.com/10.1007/BF02823296},
  copyright = {http://www.springer.com/tdm},
  langid = {english}
}

@article{Belinicher1980Photogalvanic,
  title = {The Photogalvanic Effect in Media Lacking a Center of Symmetry},
  author = {Belinicher, V I and Sturman, B I},
  year = 1980,
  month = mar,
  journal = {Soviet Physics Uspekhi},
  volume = {23},
  number = {3},
  pages = {199--223},
  issn = {0038-5670},
  doi = {10.1070/PU1980v023n03ABEH004703},
  url = {https://ufn.ru/en/articles/1980/3/b/}
}

@book{Ivchenko2012Superlattices,
  title = {Superlattices and {{Other Heterostructures}}: {{Symmetry}} and {{Optical Phenomena}}},
  shorttitle = {Superlattices and {{Other Heterostructures}}},
  author = {Ivchenko, Eougenious L. and Pikus, Grigory},
  year = 2012,
  month = dec,
  publisher = {Springer Science \& Business Media},
  isbn = {978-3-642-60650-2},
  langid = {english},
  url = {https://doi.org/10.1007/978-3-642-60650-2}
}

@article{Ma2021Topology,
  title = {Topology and Geometry under the Nonlinear Electromagnetic Spotlight},
  author = {Ma, Qiong and Grushin, Adolfo G. and Burch, Kenneth S.},
  year = 2021,
  month = dec,
  journal = {Nature Materials},
  volume = {20},
  number = {12},
  pages = {1601--1614},
  publisher = {Nature Publishing Group},
  issn = {1476-4660},
  doi = {10.1038/s41563-021-00992-7},
  url = {https://www.nature.com/articles/s41563-021-00992-7},
  copyright = {2021 Springer Nature Limited},
  langid = {english}
}

@book{Sturman2021Photovoltaic,
  title = {Photovoltaic and {{Photo-refractive Effects}} in {{Noncentrosymmetric Materials}}},
  author = {Sturman, Boris and Fridkin, Vladimir},
  year = 2021,
  month = mar,
  publisher = {Routledge},
  address = {London},
  doi = {10.1201/9780203743416},
  isbn = {978-0-203-74341-6}
}

@article{Ma2017Direct,
  title = {Direct Optical Detection of {{Weyl}} Fermion Chirality in a Topological Semimetal},
  author = {Ma, Qiong and Xu, Su-Yang and Chan, Ching-Kit and Zhang, Cheng-Long and Chang, Guoqing and Lin, Yuxuan and Xie, Weiwei and Palacios, Tom{\'a}s and Lin, Hsin and Jia, Shuang and Lee, Patrick A. and {Jarillo-Herrero}, Pablo and Gedik, Nuh},
  year = 2017,
  month = sep,
  journal = {Nature Physics},
  volume = {13},
  number = {9},
  pages = {842--847},
  publisher = {Nature Publishing Group},
  issn = {1745-2481},
  doi = {10.1038/nphys4146},
  url = {https://www.nature.com/articles/nphys4146},
  copyright = {2017 Springer Nature Limited},
  langid = {english}
}

@article{Ni2021Giant,
  title = {Giant Topological Longitudinal Circular Photo-Galvanic Effect in the Chiral Multifold Semimetal {{CoSi}}},
  author = {Ni, Zhuoliang and Wang, K. and Zhang, Y. and Pozo, O. and Xu, B. and Han, X. and Manna, K. and Paglione, J. and Felser, C. and Grushin, A. G. and {de Juan}, F. and Mele, E. J. and Wu, Liang},
  year = 2021,
  month = jan,
  journal = {Nature Communications},
  volume = {12},
  number = {1},
  pages = {154},
  publisher = {Nature Publishing Group},
  issn = {2041-1723},
  doi = {10.1038/s41467-020-20408-5},
  url = {https://www.nature.com/articles/s41467-020-20408-5},
  copyright = {2021 The Author(s)},
  langid = {english}
}

@article{Rees2020Helicitydependent,
  title = {Helicity-Dependent Photocurrents in the Chiral {{Weyl}} Semimetal {{RhSi}}},
  author = {Rees, Dylan and Manna, Kaustuv and Lu, Baozhu and Morimoto, Takahiro and Borrmann, Horst and Felser, Claudia and Moore, J. E. and Torchinsky, Darius H. and Orenstein, J.},
  year = 2020,
  month = jul,
  journal = {Science Advances},
  volume = {6},
  number = {29},
  pages = {eaba0509},
  issn = {2375-2548},
  doi = {10.1126/sciadv.aba0509},
  url = {https://www.science.org/doi/10.1126/sciadv.aba0509},
  copyright = {https://creativecommons.org/licenses/by-nc/4.0/},
  langid = {english}
}

@article{Hu2021Topological,
       author = {{Hu}, Haoyu and {Chen}, Lei and {Setty}, Chandan and {Garcia-Diez}, Mikel and {Grefe}, Sarah E. and {Prokofiev}, Andrey and {Kirchner}, Stefan and {Vergniory}, Maia G. and {Paschen}, Silke and {Cano}, Jennifer and {Si}, Qimiao},
        title = "{Topological semimetals without quasiparticles}",
      journal = {arXiv e-prints},
         year = 2021,
        month = oct,
          eid = {arXiv:2110.06182},
        pages = {arXiv:2110.06182},
          doi = {10.48550/arXiv.2110.06182},
archivePrefix = {arXiv},
       eprint = {2110.06182},
 primaryClass = {cond-mat.str-el}
}

@article{Kirschbaum2026Emergent,
  title = {Emergent Topological Semimetal from Quantum Criticality},
  author = {Kirschbaum, D. M. and Chen, L. and Zocco, D. A. and Hu, H. and Mazza, F. and Karlich, M. and Lu{\v z}nik, M. and Nguyen, D. H. and Larrea Jim{\'e}nez, J. and Strydom, A. M. and Adroja, D. and Yan, X. and Prokofiev, A. and Si, Q. and Paschen, S.},
  year = 2026,
  month = feb,
  journal = {Nature Physics},
  volume = {22},
  number = {2},
  pages = {218--224},
  issn = {1745-2473, 1745-2481},
  doi = {10.1038/s41567-025-03135-w},
  url = {https://www.nature.com/articles/s41567-025-03135-w},
  langid = {english}
}

\author{}

\setcounter{secnumdepth}{3}

\onecolumngrid
\newpage
\beginsupplement

\begin{center}
\textbf{\large Chiral Weyl--Kondo semimetal and circular photogalvanic effect in a prototype Kondo lattice system
\vspace{4pt} \\ 
SUPPLEMENTAL MATERIAL}
\end{center}

\section{Further details about the saddle point solution}\label{sec:saddle}
Consider the $f$ electrons at the energy level $\epsilon_f$ with an on-site Hubbard interaction $U$ and coupling to conduction electrons with hybridization $V$. 
We study the Kondo lattice model with Hamiltonian:
\begin{equation}\label{eq:AM}
    H = H_0 + \sum_{i,\sigma} \left[ \epsilon_f n_{f,i\sigma} + V \left( f_{i\sigma}^\dagger c_{i\sigma} + h.c. \right) \right]  \\ 
    -\mu \sum_{i,\sigma} \left( n_{c,i\sigma}+n_{f,i\sigma} \right) + \sum_i U n_{f,i\uparrow} n_{f,i\downarrow} \,,
\end{equation}
where $\sigma=\pm 1$ labels the spin degree of freedom and $H_0$ is the conduction electron Hamiltonian. 
A chemical potential $\mu$ is added to fix the total particle number. 
We refer the readers to the paragraph following Eq.~(1) in the main text for the definitions of the parameters and operators.
In order to study the symmetry-enabled Kondo screened phase, we consider the soluble case realized in the infinite $U$ limit; here, the model is solved at the saddle-point level, which formally develops in a large-$N$ limit and in terms of an auxiliary-boson representation~\cite{Hewson1993Kondo}. 
The effective Hamiltonian is given as follows:
\begin{equation}
\label{eqn:model}
    H = H_0 + \sum_{i,\sigma} \left[ \epsilon_f n_{f,i\sigma} + rV \left( f_{i\sigma}^\dagger c_{i\sigma} + h.c. \right) \right]  \\ 
    -\mu \sum_{i,\sigma} \left( n_{c,i\sigma}+n_{f,i\sigma} \right) + \lambda \sum_i \left( n_{f,i} +r^2-Q \right) \, .
\end{equation}
Here, $r$ is the condensed auxiliary-boson field $\left\langle b_i \right\rangle$, while $c_{i\sigma}$ and $f_{i\sigma}$ represent the conduction and heavy electrons at site $i$ with spin $\sigma$ while $n_{f,i}=\sum_{\sigma}f^\dagger_{i\sigma}f_{i\sigma}$ represents the number of $f$ electrons at site $i$.
The $\lambda$ term comes from the Lagrange multiplier of the local filling constraint $n_{f,i}+b_i^\dagger b_i=Q$.
The self-consistent equations read as follows:
\begin{align}
    \langle n_{f,i}\rangle + r^2 &= Q \,, \label{eqn:self1} \\
    V \sum_{\sigma}\langle f_{i\sigma}^\dagger c_{i\sigma}+ h.c\rangle   &= -2\lambda r \,, \label{eqn:self2} \\ 
    \frac{1}{N_{u.c.}} \sum_{i} \left( \langle n_{f,i}\rangle +\langle n_{c,i}\rangle \right) &= \nu \,, \label{eqn:self3} 
\end{align}
where $\langle\cdots\rangle$ means the expectation value, $h.c.$ means hermitian conjugate, $n_{c,i} = \sum_{\sigma} c_{i,\sigma}^{\dagger} c_{i,\sigma}$, $N_{u.c.}$ is number of unit cells and $\nu$ is filling per unit cell.
The self-consistent equations are solved by iteration to determine parameters $\lambda$, $r$, and $\mu$.

Here the symmetric conduction-electron Hamiltonian $H_0$ is constructed to satisfy symmetries of space group no.~78 ($P4_3$), time-reversal ($\mathcal{T}$) symmetry, and Hermitian condition~\cite{cano2021band}
\begin{align}
    H_0 &= \frac{1}{|\mathcal{C}|}\sum_{g\in \mathcal{C}} g H_0 g^{-1} \,,\quad H_0 = \frac12 \left( H_0 + \mathcal{T} H_0 \mathcal{T}^{-1} \right)  \,, \quad 
    H_0 = \frac12 \left( H_0 + H_0^\dagger \right) \,. 
\end{align}
Here $\mathcal{C}=\{ E, \{ C_{4z} | 0,0,\frac{3}{4} \}, \{ C_{2z} | 0,0,\frac{1}{2} \}, \{ C_{4z}^3 | 0,0,\frac{1}{4} \}   \}$ where $E$ is identity and $\{ C_{4z} | 0,0,\frac{3}{4} \} $ is the screw symmetry. 
By examining these constraints, we can derive the symmetric hopping terms including SOC terms for real space tight-binding models.
Here we only include nearest and next-nearest neighbor terms for the prototype model. 

Throughout this work, we choose bare parameters $\epsilon_f=-15 t$, $V=6 t$, $Q=1$ and $\nu=4$ for the prototype model where $t$ is the $c$ electron nearest neighbor hopping amplitude.
The resulting self-consistent parameters are $\lambda=5.691t$, $r=0.206$ and $\mu=-9.473t$, calculated over a $32\times32\times32$ lattice with error tolerance $1\times 10^{-5}$.

\section{Further details about symmetry-enforced band crossings}\label{sec:symmetry}

In this section, we explain how the symmetries of space group no.~78 ($P4_3$) enforce band crossings.
We denote the Bravais lattice vectors as $\mathbf{a}_i$ ($i=1,2,3$), and we use the notation $\{ R | n_1, n_2, n_3 \}$ to denote a space-group operation, where $R$ is a point group operation and $(n_1 , n_2 , n_3)$ denotes a translation of $\sum_{i=1}^{3} n_i \mathbf{a}_i$.
Throughout this section, the presence of SOC and time-reversal ($\mathcal{T}$) symmetry are assumed.
A crystal momentum $\mathbf{k}$ is denoted as $(k_x , k_y , k_z)$ in the Cartesian coordinate.
We work in the unit that the Brillouin zone is defined by the Cartesian components $k_x$, $k_y$, and $k_z$ being in the range from $-\pi$ to $\pi$.
The high-symmetry points are $\Gamma = (0,0,0)$, $X = (\pi,0,0)$, $M = (\pi,\pi,0)$, $Z = (0,0,\pi)$, $R = (\pi,0,\pi)$, and $A = (\pi,\pi,\pi)$.
We use $\mathbf{k}_1 - \mathbf{k}_2$ to denote a line starting at $\mathbf{k}_1$ and ending at $\mathbf{k}_2$.
In the following, we focus on how $\{ C_{4z} | 0,0,\frac{3}{4} \}$, $\{ C_{2z} | 0,0,\frac{1}{2} \}$, and $\{ C_{2z} \mathcal{T} | 0,0,\frac{1}{2} \}$ lead to symmetry-enforced band crossings and degeneracies~\cite{hirschmann2021symmetryenforced}.

Both $\Gamma-Z$ and $M-A$ are invariant upon $\{ C_{4z} | 0,0,\frac{3}{4} \}$.
Hence, the Bloch eigenstate along $\Gamma-Z$ and $M-A$ can be labeled by the $\{ C_{4z} | 0,0,\frac{3}{4} \}$ eigenvalue.
Specifically, as $\Gamma-Z$ and $M-A$ are parametrized by the crystal momentum component $k_z$, the $\{ C_{4z} | 0,0,\frac{3}{4} \}$ eigenvalue of the corresponding Bloch eigenstate can be $e^{\frac{-i3k_z}{4}} e^{\frac{i\pi}{4}}$, $e^{\frac{-i3k_z}{4}} e^{\frac{i3\pi}{4}}$, $e^{\frac{-i3k_z}{4}} e^{\frac{i5\pi}{4}}$, or $e^{\frac{-i3k_z}{4}} e^{\frac{i7\pi}{4}}$.
Using these $\{ C_{4z} | 0,0,\frac{3}{4} \}$ eigenvalues, the band connectivity along $\Gamma-Z$ and $M-A$ can be constructed, leading to the accordion-type band connectivity in Fig.~2 and Fig.~3(c) in the main text.

$X-R$ is invariant upon $\{ C_{2z} | 0,0,\frac{1}{2} \}$.
Hence, the Bloch eigenstate along $X-R$ can be labeled by the $\{ C_{2z} | 0,0,\frac{1}{2} \}$ eigenvalue.
Specifically, as $X-R$ are parametrized by the crystal momentum component $k_z$, the $\{ C_{2z} | 0,0,\frac{1}{2} \}$ eigenvalue of the corresponding Bloch eigenstate can be $e^{\frac{-ik_z}{2}}e^{\frac{i\pi}{2}}$ or $e^{\frac{-ik_z}{2}}e^{\frac{i3\pi}{2}}$.
Using these $\{ C_{2z} | 0,0,\frac{1}{2} \}$ eigenvalues, the band connectivity along $X-R$ can be constructed, leading to the hourglass-type band connectivity in Fig.~2 and Fig.~3(c) in the main text.

For any Bloch eigenstate with a crystal momentum $\mathbf{k} = (k_x , k_y , \pi)$, the action of $\{ C_{2z} \mathcal{T} | 0,0,\frac{1}{2} \}$ on it transforms it into a Bloch eigenstate with the same crystal momentum.
We note that $\{ C_{2z} \mathcal{T} | 0,0,\frac{1}{2} \}^2 = \{ E | 0,0,1 \}$ where $E$ is the identity.
Therefore, for any Bloch eigenstate with $\mathbf{k} = (k_x , k_y , \pi)$, $\{ C_{2z} \mathcal{T} | 0,0,\frac{1}{2} \}^2$ acts on it as $-1$.
Furthermore, $\{ C_{2z} \mathcal{T} | 0,0,\frac{1}{2} \}$ is an anti-unitary operation.
Therefore, Kramers' theorem enforces a twofold degeneracy in the band structure on the $k_z = \pi$ plane.
This can be seen along $Z - R - A - Z$ in Fig.~2 in the main text.

As a final remark, due to the enforced twofold degeneracy on the $k_z = \pi$ plane, the Kramers--Weyl points from the structural chirality~\cite{Chang2018Topological} can only appear at the time-reversal invariant momenta (TRIMs) on the $k_z = 0$ plane, namely $\Gamma$, $X$, and $M$.

\section{Further details about chiral charge calculations}\label{sec:charge}

The chiral charge of a Weyl point is the quantized total flux of the Berry curvature threading through a closed surface enclosing the Weyl point~\cite{Armitage2018Weyl}.
We start by numerically locating the position of the Weyl point.
We then enclose the Weyl point by a cube with a suitable size.
We divide each surface of the cube into small plaquettes with a suitable density.
We next evaluate the Berry phase of the chosen band along the boundary of each small plaquette.
The path along which the Berry phase is evaluated is the right-handed path determined by the surface normal vector of the small plaquette pointing outwards from the cube.
Summing the Berry phases of all small plaquettes yields the chiral charge of the Weyl point.
We use this method to obtain the chiral charges of the Weyl points labeled in Fig.~3(c) in the main text.

\section{Further details about the CPGE calculations}\label{sec:CPGE}
CPGE is a second order nonlinear optical effect that generates a current response to circularly polarized light. 
To be concrete, we consider two cases: alternating current (AC) CPGE and direct current (DC) CPGE.
AC CPGE is a current of frequency $\Omega = \omega_1 - \omega_2$ generated by lissajous-polarized light of two perpendicular polarizations with frequencies $\omega_1$ and $\omega_2$. 
The DC CPGE is a DC current generated by a single circularly polarized light of frequency $\omega$. 
The conductance of AC CPGE is defined as~\cite{DeJuan2020Difference} 
\begin{equation}
    \frac{dj^a}{dt} = \beta^{aa'} \left( \mathbf E(\omega_1) \times  \mathbf E^* (-\omega_2)\right)_{a'} \,,
\end{equation}
where $a$, $a'$ are spatial indices, $\mathbf E(\omega)$ is the electric field of frequency $\omega$ and $\beta^{aa'}$ is the CPGE conductance tensor. 
Here $dj^a/dt=i\Omega j^a$ is the $\Omega$-frequency response.  
When $\omega_1=\omega_2=\omega$, the relaxation time effect cannot be ignored, and we get a DC current at zero frequency. 
The conductance of DC CPGE is defined as~\cite{DeJuan2017Quantized}
\begin{equation}
    \frac{j^a}{\tau} = \beta^{aa'} \left( \mathbf E(\omega) \times  \mathbf E^* (-\omega)\right)_{a'} \,.
\end{equation}
where $\tau$ is the relaxation time. Both AC and DC CPGE share the same response tensor.

It has been shown that in noninteracting Weyl semimetals, the trace of CPGE conductance tensor $\Tr[\beta]$ is quantized to $C\frac{\pi e^3}{h^2}$ in certain frequency window where $C$ is the monopole charge of the Weyl point~\cite{DeJuan2017Quantized}.
In chiral systems, the contributions to the CPGE from Weyl points of opposite chirality cancel each other out. Therefore, quantized CPGE can only be observed in chiral Weyl semimetals, where the Weyl points of opposite chirality are at different energies.
Weyl--Kondo semimetals (WKSMs) are a class of strongly correlated topological materials, where the heavy fermion Weyl nodes emerge from the hybridization between conduction electrons Weyl fermions and the local moments of localized $f$ electrons. 
These systems exhibit both nontrivial topology and strong interactions, making them ideal platforms to explore the interplay between correlation effects and topological responses.

\begin{figure}[t]
    \centering
    \includegraphics[width=\linewidth]{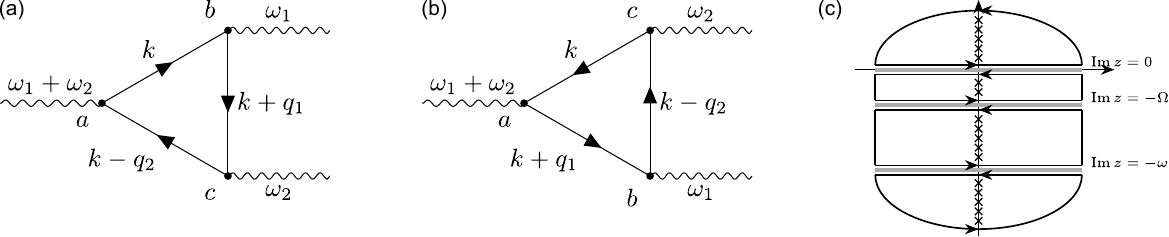}
    \caption{
    (a) and (b) are the triangle diagrams that determine chiral anomaly and CPGE of the single-Weyl node models. 
    (c) Branch cut and deformation of contour integral.  
    }
    \label{fig:chiral_anomaly}
\end{figure}

CPGE is determined by the two triangle diagrams shown in Fig.~\ref{fig:chiral_anomaly}(a) and (b):
\begin{equation}
    j^a(\Omega) = \frac{1}{\omega_1\omega_2}\left[ \chi^{abc}(\omega_1,\omega_2)+\chi^{acb}(\omega_2,\omega_1)\right] E^b(\omega_1)E^c(\omega_2)
\end{equation}
where 
\begin{equation}
\label{eqn:chi_integral}
    \chi^{abc}(i\omega_1,i\omega_2) = T\sum_{\omega_n} \int \frac{d^3k}{(2\pi)^3} \Tr \left( \gamma^a G(i\omega_n+i\omega_1,\mathbf k) \gamma^b G(i\omega_n+i\Omega,\mathbf k) \gamma^c G(i\omega_n,\mathbf k) \right) \,,
\end{equation}
where $T$ is the temperature, $\omega_n=2\pi T (n+\frac12)$ is the fermionic Matsubara frequency, $\Omega=\omega_1+\omega_2$, $\Omega\rightarrow 0$, $\gamma^a$ is the vertex matrix and $G$ is the Green's function matrix. 
Then we can take $T\sum_{i\omega_n} \dots = \frac{1}{2\pi i}\oint_C dz f(z) \dots$, where $f$ is Fermi--Dirac distribution. 
Note diamagnetic terms vanish in the models we consider. 
Considering deforming the contour integral to that is shown in Fig.~\ref{fig:chiral_anomaly}(c), both AC and DC CPGE coefficient $\beta^{aa'}$ are compactly determined by the Sipe--Shkrebtii formula~\cite{Sipe2000Secondordera}:
\begin{align}
    \beta^{aa'} = \pi\epsilon^{a'bc} \int_{\mathbf k} \sum_{nm} \frac{\partial \epsilon_{nm}}{\partial k_a} r^b_{nm} r^c_{mn} f_{nm} \delta(\omega-\epsilon_{nm}) 
\end{align}
where $\int_{\mathbf{k}}\equiv \int \frac{d^3 k}{(2\pi)^3}$, $\epsilon_{nm}=\epsilon_n(\mathbf{k})-\epsilon_m(\mathbf{k})$, $f_{nm}=f_n-f_m$, and $r^b_{nm}=i\langle u_n(\mathbf{k})|\partial_{k_b}|u_m(\mathbf{k})\rangle$. Here $\epsilon_n(\mathbf{k})$ and $|u_n(\mathbf{k})\rangle$ are the energies and the periodic part of Bloch states, $f_n$ is the Fermi--Dirac distribution function for the $n$-th band.
This is a second order optical response involving only bands at resonance $\omega=\epsilon_{nm}$.  

\subsection{CPGE of a single Weyl point}
In this subsection we review the quantization of CPGE coefficient for a single Weyl point in a chiral Weyl semimetal.
Without loss of generality, we only consider one $+1$ chirality Weyl point and it can be easily generalized to $-1$ chirality Weyl point. 

Consider a noninteracting Weyl semimetal Hamiltonian coupling to an external electromagnetic field
\begin{equation}\label{eqn:Weyl_Hamiltonian_sm}
    H_0=\sum_{\mathbf k, \alpha,\alpha'} c^\dagger_{{\mathbf k},\alpha}  \left[ v_{\mathrm{F}} ({\mathbf k}+{\mathbf A}) \cdot \vec{\sigma}_{\alpha\alpha'} - \mu \right] c_{{\mathbf k},\alpha'} 
\end{equation}
where $\mathbf A$ is the vector potential, $v_{\mathrm{F}}$ is the Fermi velocity, $\vec \sigma$ are the Pauli matrices and $\mu$ is the chemical potential.
For this chiral model, the current operator is identical to the chiral current operator, which is $\hat{j}=\sum_{\mathbf k, \alpha,\alpha'} c^\dagger_{{\mathbf k},\alpha}  v_{\mathrm{F}} \vec{\sigma}_{\alpha\alpha'} c_{{\mathbf k},\alpha'} $. 
Due to linearity in $k$, this model does not have diamagnetic term. 
As a result, the second order optical response is given by the two triangle diagrams, which is exactly the chiral anomaly~\cite{Adler1969AxialVector,Bell1969PCAC,Adler1969Absence}.

Since for two level systems the projectors of conduction/valence band are:
\begin{equation}
    P_\xi(\mathbf k)= \frac{I + \xi \hat{\mathbf k}\cdot \vec \sigma}{2},
\end{equation}
where $\xi=\pm$ labels conduction/valence band and $\hat{\mathbf k}=\mathbf k/|\mathbf k|$ is the unit vector in the direction of $\mathbf k$.
Then the Green's function matrix can be expressed as
\begin{equation}
    G(i\omega_n,{\mathbf k}) = \frac{P_+}{i\omega_n-v_{\mathrm{F}} k + \mu} + \frac{P_-}{i\omega_n+v_{\mathrm{F}} k + \mu}
\end{equation}
where $k=|\mathbf k|$ is the magnitude of the momentum $\mathbf k$.
Plugging the Green's function into the triangle diagrams (note there is no diamagnetic term in this $k$-linear model), we get
\begin{align}
    j^a(\Omega) &=-\frac{2\pi i}{\Omega}  \int_{\mathbf k}  \left( \Tr \left[ \frac{\partial H_0}{\partial k_a}  P_+ \frac{\partial H_0}{\partial k_b} P_- \frac{\partial H_0}{\partial k_c} P_+\right] + \Tr \left[ \frac{\partial H_0}{\partial k_a}  P_- \frac{\partial H_0}{\partial k_b} P_+ \frac{\partial H_0}{\partial k_c} P_-\right] \right) f_{+-} \delta(\omega-\epsilon_{+-})  A_b A_c \\
    &= \frac{2\pi }{\Omega} \frac{4\pi}{3} \frac{v_{\mathrm{F}}^3}{(2\pi)^3} \epsilon^{abc}  \left(\frac{\omega}{2 v_{\mathrm{F}}}\right)^2 \Theta(\omega-2|\mu|) \frac{1}{2 v_{\mathrm{F}}} A_bA_c \times 2\\ 
    &=\frac{1}{12\pi}\frac{1}{\Omega} \epsilon^{abc}\Theta(\omega-2|\mu|) E^b(\omega+\Omega)E^c(\omega)\label{eqn:CPGE}
\end{align}
where Pauli algebra is used. 
Adding physical constants back, the trace of the coefficient tensor is
\begin{align}
    \beta_0 &= \frac{\pi e^3}{h^2}
\end{align}
For a single Weyl point, the photocurrent conductivity is quantized for photon energy $|\omega|>2|\mu|$. 
It can be proved that for generic noninteracting two-band model with Weyl point of chiral charge $C$, the trace of the coefficient tensor is quantized to $\beta=C\beta_0$. If the model has multiple Weyl points, the response is proportional to the total charge surrounded by the resonant surface at frequency $\omega$.

\subsection{CPGE of a single Weyl--Kondo point}
Here we consider the model of a single Weyl--Kondo point:
\begin{equation}
\label{eqn:model_single}
    H=H_0+\sum_{i,\alpha} \left( \epsilon_f^* f_{i\alpha}^\dagger f_{i\alpha} + V^* \left[c_{i\alpha}^{\dagger} f_{i\alpha}+h.c.\right] \right)  \,,
\end{equation}
where $H_0$ is given by Eq.~\eqref{eqn:Weyl_Hamiltonian_sm}. 
For convenience, we define the renormalized hybridization $V^*=r V$ and the renormalized $f$ electron energy $\epsilon_f^*=\epsilon_f+\lambda$.
In this subsection, we study the CPGE coefficient in this effective model. 

The effective Hamiltonian Eq.~\eqref{eqn:model_single} have four bands in momentum space.
We adopt notation $(\xi\eta)=(\pm \pm)$ to label their energies:
\begin{equation}
    \epsilon_{\xi\eta}(\mathbf k) = \frac{\xi v_{\mathrm{F}} k+\epsilon_f^*-\mu}{2} +\eta \sqrt{ \left( \frac{\xi v_{\mathrm{F}} k-\epsilon_f^*-\mu}{2} \right)^2+V^{*2} }  \,,
\end{equation}
where $k =|\mathbf k|$ is the magnitude of the momentum $\mathbf k$.
A schematic picture of the band structure is shown in Fig.~3(a).
$\epsilon_{++}$ and $\epsilon_{-+}$ form the $c$ electron Weyl point, while $\epsilon_{+-}$ and $\epsilon_{--}$ form the $f$ electron Weyl point. 
We first assume Fermi energy is at the $f$ electron Weyl point for simplicity. In later sections we will use self-consistent mean field theory to determine the accurate Fermi energy.

Let us rewrite the mean field Hamiltonian in the basis $\Psi^\dagger_k = (c^\dagger_k,~f_k^\dagger)$:
\begin{align}
    H &= \sum_k
    \begin{pmatrix}
    c^\dagger_{k,\alpha}~,~f_{k,\beta}^\dagger 
    \end{pmatrix}
    \begin{pmatrix}
        v_{\mathrm{F}} ({\mathbf k}+{\mathbf A}) \cdot \vec{\sigma}_{\alpha\alpha'} - \mu \delta_{\alpha\alpha'} & V^* \delta_{\alpha \beta'}\\
        V^* \delta_{\alpha' \beta} & \epsilon_f^* \delta_{\beta \beta'}
    \end{pmatrix}
    \begin{pmatrix}
        c_{k,\alpha'}\\ f_{k,\beta'}
    \end{pmatrix} \n \\
    &\equiv \sum_k \Psi^\dagger_k H_k \Psi_k 
\end{align}

The Green's function $G$ of the mean field Hamiltonian is:
\begin{align}
    G(z,k) \equiv \begin{pmatrix}
        G_{cc} & G_{cf} \\
        G_{fc} & G_{ff}
    \end{pmatrix} &=\frac{1}{z - H_k}
\end{align}
where $G_{cc}$ is 
\begin{align}
    G_{cc} &= \frac{(z-\epsilon_f)P_+}{(z-\epsilon_{++})(z-\epsilon_{+-})} +  \frac{(z-\epsilon_f)P_-}{(z-\epsilon_{-+})(z-\epsilon_{--})} \\
    &= \frac{P_+}{z - v_{\mathrm{F}} k + \mu -\Sigma(z)} +  \frac{P_-}{z + v_{\mathrm{F}} k + \mu-\Sigma(z)} \\
    &= \sum_{\xi\eta} \frac{Z_{\xi\eta} P_\xi }{z - \epsilon_{\xi\eta}} \,, \label{eqn:Gcc}
\end{align}
and the self-energy for $G_{cc}$ is
\begin{equation}
    \Sigma(z)=\frac{V^{*2}}{z-\epsilon_f^*} \,.
\end{equation}
The quasiparticle weight $Z_{\xi\eta}$ for each pole at momentum $\mathbf k$ is defined as
\begin{align}
    Z_{\xi\eta}& = (1-\partial_z\Sigma\big|_{z=\epsilon_{\xi\eta}})^{-1} \\
    &= \left( 1 + \frac{V^{*2}}{(z-\epsilon_f^*)^2} \right)^{-1} \Bigg|_{z=\epsilon_{\xi\eta}} \,.
\end{align}
It is evident that when $V=0$ it reduced to the Green's function without hybridization.
The physical meaning of $Z_{\xi\eta}$ is the conduction electron weight of the band labeled by $(\xi,\eta)$. This can be seen from the form of retarded/advanced Green's function near the poles:
$[G_{cc}^{R/A}(\omega)]^{-1}\approx Z_{\xi\eta} \left(\omega - \epsilon_{\xi\eta} \pm i 0^+\right)$
where the spectral function integral around the pole gives conduction electron weight $Z_{\xi\eta}$. 

The conventional definition of quasiparticle weight of heavy fermion quasiparticles in parton formalism is given by the self-consistent parameter $r^2=1-\langle n_{fi}\rangle$. Thus, we have 
\begin{equation}
    r^2 = 1-\sum_{\xi,\eta=\pm} \int_{\mathbf{k}} (1-Z_{\xi \eta}) f_{\xi\eta}
\end{equation}
where $f$ is Fermi--Dirac distribution. When the $f$-Weyl point exactly sits at the Fermi energy, we have a simple relation $r^2=Z_{--}$.

Similarly, we have the Green's function $G_{ff}$:
\begin{align}\label{eqn:Gff}
    G_{ff} &= \frac{P_+}{z - \epsilon_f^* -\Sigma_{f+}(z)} +  \frac{P_-}{z - \epsilon_f^* -\Sigma_{f-}(z)} \,,
\end{align}
where the self-energies $\Sigma_{f\pm}(z)$ are given by
\begin{equation}
    \Sigma_{f\pm}(z) = \frac{V^2}{z \mp v_{\mathrm{F}} k + \mu} \,.
\end{equation}

\subsubsection{CPGE at $f$ electron Weyl point}
Now we consider the CPGE in WKSMs. 
In this subsection, we will focus on the $f$ electron Weyl point.
The current operator of the mean field theory coupling to the vector potential is
\begin{align}
    \hat{j}^a &= e\frac{\delta H}{\delta A_a} = \sum_k \begin{pmatrix}
    c^\dagger_{k,\alpha}~,~f_{k,\beta}^\dagger 
    \end{pmatrix}
    \begin{pmatrix}
        ev_{\mathrm{F}}  \sigma^a_{\alpha\alpha'} & \n \\
         & 0
    \end{pmatrix}
    \begin{pmatrix}
        c_{k,\alpha'}\\ f_{k,\beta'}
    \end{pmatrix} \,.
\end{align}
Therefore, only $G_{cc}$ contributes to the CPGE response.

We focus on the case where $\mu<0$ and the Fermi energy is at the $f$ electron Weyl point, i.e., $E_f=\epsilon_{--}^0$. 
In this case, the angle integral is the same as the Weyl semimetal case while the relevant $k$-integral has one possible contribution from $P_-P_+P_+$ term while the $P_-P_-P_-$ term vanishes after integration over angles.

We use the Green's function Eq.~(\ref{eqn:Gcc}) to evaluate the integral for $P_-P_+P_+$ term for CPGE.
The result is:
\begin{equation}
    \beta(\omega) = \beta_0 \sum_{\eta=\pm} Z_{--}Z_{+\eta} \left(\frac{2 v_{\mathrm{F}} k^*_\eta}{\omega}\right)^2 \Theta(k^*_\eta-k_{\mathrm{F}}) \,.
\end{equation}
Here $k_{\mathrm{F}}$ is the momentum at Fermi energy and the renormalized resonant momentum $k^*_\eta$ is the solution of 
\begin{equation}\label{eqn:k_asterisk_f}
    \omega = \epsilon_{+\eta}(k^*_\eta) - \epsilon_{--}(k^*_\eta) \,.
\end{equation}
There is a critical frequency
\begin{align}\label{eqn:omegac}
    \omega_c &= \sqrt{(\epsilon_f^*+\mu)^2+4V^{*2}} \,.
\end{align}
If $\omega<\omega_c$ there is only one positive solution $k^*$ of Eq.~(\ref{eqn:k_asterisk_f}) for the process between $\epsilon_{--}^*$ and $\epsilon_{+-}^*$.
If $\omega>\omega_c$ there are two positive solutions, where the second solution corresponds to the process between $\epsilon_{--}^*$ and $\epsilon_{++}^*$.

Now we analyze how $\beta(\omega)$ behaves at $\omega \rightarrow 0$ and $\omega \rightarrow \infty$.
In the small $\omega$ limit, there is only one $k^*\approx 0$ for $\beta=-$.
We have $(2k^*_-Z_{--}/\omega)|_{k=0}=(2k^*_-Z_{+-}/\omega)|_{k=0}=1$. Therefore, $\beta(\omega\rightarrow 0) \rightarrow \beta_0$.

In the large $\omega$ limit, $Z_{--}|_{k=\infty}=Z_{++}|_{k=\infty}=1$, $Z_{+-}|_{k=\infty}=Z_{-+}|_{k=\infty}=0$, $k^*_+=2\omega$.
Therefore, both $\beta(\omega\rightarrow \infty) \rightarrow \beta_0$.

\subsubsection{CPGE at $c$ electron Weyl point}
In this subsection, we consider the CPGE at the $c$ electron Weyl point. 
We focus on the case where $\mu>0$ and the Fermi energy is above the $f$ electron Weyl point, i.e., $\epsilon_f=\epsilon_{++}(0)$. 
In this case, the angle integral is the same as the Weyl semimetal case while the relevant $k$-integral has two possible contributions: $P_+P_-P_-$ and $P_+P_+P_+$.

We use the Green's function Eq.~(\ref{eqn:Gcc}) to evaluate the integral for $P_-P_+P_+$ term for CPGE.

The CPGE coefficient is:
\begin{equation}
    \beta(\omega) = \beta_0 \sum_{\eta=\pm} Z_{++}Z_{-\eta} \left(\frac{2k^*_\eta}{\omega}\right)^2 \Theta(k^*_\eta-k_{\mathrm{F}}) \,.
\end{equation}
Here the resonant momentum $k^*_\eta$ is the solution of 
\begin{equation}\label{eqn:k_asterisk_c}
    \omega = \epsilon_{++}(k^*_\eta) - \epsilon_{-\eta}(k^*_\eta) \,.
\end{equation}
The critical frequency is again $\omega_c$ in Eq.~(\ref{eqn:omegac}). 
We can show that both coefficients have properties: $\beta(\omega\rightarrow 0)=\beta_0$ and $\beta(\omega\rightarrow \infty)=\beta_0$.

\end{document}